\documentclass[aps,11pt,secnumarabic,showpacs,showkeys,amsmath,amssymb,groupedaddress,%
eqsecnum,a4paper,tightenlines,preprintnumbers]{revtex4}

\newcommand{\Tr}{\operatorname{Tr}}
\newcommand{\rmd}{\mathrm{d}}
\newcommand{\rmi}{\mathrm{i}}

\newcommand{\rmq}{\mathrm{q}}
\newcommand{\rmc}{\mathrm{c}}
\newcommand{\rmf}{\mathrm{f}}
\begin{document}
\preprint{Politecnico di Milano, Math.\ Dept., Quaderno n.604/P}

\title{Instruments and channels in quantum information theory}

\author{Alberto Barchielli}
\altaffiliation[Also: ]{Istituto Nazionale di Fisica Nucleare, Sezione di Milano}
\email{Alberto.Barchielli@polimi.it} \homepage[Home page: link to ``Personale'' in
]{http://www.mate.polimi.it}
\affiliation{Politecnico di Milano, Dipartimento di Matematica, \\
Piazza Leonardo da Vinci 32, I-20133 Milano, Italy.}
\author{Giancarlo Lupieri}
\altaffiliation[Also: ]{Istituto Nazionale di Fisica Nucleare, Sezione di Milano}
\email{Giancarlo.Lupieri@mi.infn.it}
\affiliation{Universit\`a degli Studi di Milano,
Dipartimento di Fisica,\\ Via Celoria 16, I-20133 Milano, Italy.}

\date{August 2004}

\begin{abstract}
While a POV measure gives the probabilities in a quantum measurement, an instrument gives both
the probabilities and the a posteriori states. By interpreting the instrument as a quantum
channel and by using the typical inequalities for the quantum and classical relative
entropies, many bounds on the classical information extracted in a quantum measurement, of the
type of Holevo's bound, are obtained in a unified manner.
\end{abstract}



\maketitle

\section{\label{intro}Introduction}

Let us consider a quantum system QS living in a Hilbert space $\mathcal{H}$; the (normal)
states of QS are represented by \emph{statistical operators} or \emph{density matrices}. A
problem which appears in the field of quantum communication and in quantum statistics is the
following: a collection of statistical operators, with some a priori probabilities, describes
the possible initial states of the system and an observer wants to decide by means of a
quantum measurement on QS in which of these states the system is. The quantity of information
given by the measurement is the classical mutual information $I_\rmc$ of the input/output
joint distribution (Shannon information); interesting upper and lower bounds for $I_\rmc$, due
to the quantum measurement, are given in the literature
\cite{Hol73,YueO93,Scu95,SchWW96,Hal97,Jac03}. Usually the measurement is described by a
\emph{generalized observable} or \emph{positive operator valued} (POV) \emph{measure}; an
exception is the paper \cite{SchWW96}, which considers also the information left in the
post-measurement states.

With respect to a POV measure, a more detailed level of description of the quantum measurement
is represented by a different mathematical object, the \emph{instrument} \cite{Dav76,Oza84}:
given a state (the preparation) as input, it gives as output not only the probabilities of the
outcomes but also the state after the measurement, conditioned on the observed outcome (the a
posteriori state). We can think the instrument to be a channel: from a quantum state (the
pre-measurement state) to a quantum/classical state (a posteriori state plus probabilities).
The mathematical formalization of the idea that an instrument \emph{is} a channel is central
in our paper and allows for a unified approach to various bounds for $I_\rmc$ and for related
quantities, which derive from the description of the quantum measurement with a well precise
instrument. The most interesting inequality is a strengthening of Holevo's bound; in the
finite case it has been obtained in Ref.\ \cite{SchWW96} where the authors introduce a
specific model of the measuring process (without speaking explicitly of intruments) and use
the strong subadditivity of the von Neumann entropy. The introduction of the general notion of
instrument, the association to it of a channel and the use of Uhlmann's monotonicity theorem
allows us to obtain the same result in a more direct way and to extend it to a more general
set up.

To maintain things at a sufficiently simple mathematical level, in Sections
\ref{sec:instrument} and \ref{sec:Hol} we shall develop and present all results in the case of
a finite-dimensional Hilbert space, a finite alphabet and an instrument with finite outcomes;
in the last section we shall just present the results in the infinite case. In Section
\ref{sec:instrument} we introduce the notion of instrument and we show how to associate a
channel to it; some inequalities on various relative entropies are deduced from Ulhmann's
theorem. From such inequalities we obtain in Section \ref{sec:Hol} the bounds on the quantity
of information which can be extracted by using an instrument as decoding apparatus. From the
main bound we obtain in a straight way also a result by Groenewold, Lindblad, Ozawa
\cite{Gro71,Lin72,Oza86} on the positivity of the \emph{quantum information gain} given by an
instrument. In Section \ref{sec:instrgen} we give a quick presentation of the main results in
the general case: an infinite dimensional Hilbert space, a general instrument with any kind of
outcomes, a generic alphabet, even a continuous one as already considered in \cite{YueO93}.

Some of the informational quantities presented here have been studied in \cite{Bar01,BarL04}
in the case of instruments describing continual measurements. The definition of quantum
relative entropy and its properties (in particular Uhlmann's theorem) can be found in the book
by Ohya and Petz \cite{OhyP93}, whose Part I is dedicated to the finite-dimensional case,
while the rest of the book treats the general case.

\section{\label{sec:instrument}Instruments and channels, finite case}
Let us start by considering a finite-dimensional complex Hilbert space $\mathcal{H}
=\mathbb{C}^n$ and an instrument with finitely many outcomes. Let us denote by $M_n$ the
algebra of the complex ($n\! \times\! n$)-matrices and by $\mathcal{S}_n\subset M_n$ the set
of statistical operators on $\mathbb{C}^n$.

\subsection{\label{subsec:finiteInst}Instruments, probabilities and a posteriori states}
In this finite context, any instrument $\mathcal{I}$ is representable as
\begin{subequations}
\begin{gather}
\mathcal{I}(F)[\rho]= \sum_{\omega \in F} \mathcal{V}(\omega)[\rho], \qquad \forall F\subset
\Omega, \quad \forall \rho\in M_n,
\\
\mathcal{V}(\omega)[\rho] = \sum_{k\in K} V_k^\omega \rho V_k^{\omega \dagger} \,,
\\
\sum_{\omega \in\Omega} E(\omega)=\openone, \qquad E(\omega)=\sum_{k\in K}
 V_k^{\omega \dagger}V_k^\omega,
\end{gather}
\end{subequations}
where $\Omega$ is the \emph{value space} (the finite set of possible outcomes), $V_k^\omega\in
M_n$ and $K$ is a suitable finite set. Note that $E$ is a POV measure, the POV measure
associated with $\mathcal{I}$. If the pre-measurement state is $\rho\in \mathcal{S}_n$, the
probability of the result $\{\omega\in F\}$, $F\subset \Omega$, is
\begin{equation}\label{probI}
P_\rho(F)=\sum_{\omega \in F} p_\rho(\omega)=  \Tr\{\mathcal{I}(F)[\rho]\}, \qquad
p_\rho(\omega)=\Tr\{E(\omega) \rho\} =\Tr\{\mathcal{V}(\omega)[\rho]\},
\end{equation}
and the post-measurement state, conditioned on this result, is
$\frac{\mathcal{I}(F)[\rho]}{\Tr\{\mathcal{I}(F)[\rho]\}}$. When $F$ shrinks to a single
point, the conditional post-measurement state reduces to what is called the \emph{a posteriori
state}
\begin{equation}\label{apostI}
\hat\rho(\omega;\rho)=\frac{\mathcal{V}(\omega)[\rho]}{p_\rho(\omega)}\qquad \text{if } \
p_\rho(\omega)>0\,;
\end{equation}
this definition has to be completed by defining arbitrarily $\hat\rho(\omega;\rho)$ for the
points $\omega$ for which $p_\rho(\omega)=0$. The a posteriori state is the state to be
attributed to the quantum system after the measurement when we know that the result of the
measurement has been exactly $\{\omega\}$. On the opposite side, we have the unconditional
post-measurement state or \emph{a priori state}
\begin{equation}\label{apriori}
\mathcal{I}(\Omega)[\rho]= \sum_{\omega \in \Omega} \mathcal{V}(\omega)[\rho]\,;
\end{equation}
it is the state to be attributed to the system after the measurement, when the result is not
known.

\subsection{\label{subsec:sec}States, entropies, channels}
\subsubsection{Algebras and states}
To formalize the idea that an instrument is a channel, we need to introduce the spaces
$\mathcal{C}(\Omega; M_n)$ of the functions from $\Omega$ into $M_n$ and
$\mathcal{C}(\Omega)\equiv \mathcal{C}(\Omega; M_1)$, which are finite $C^*$-algebras, as
$M_n$. A state on a finite $C^*$-algebra is a normalized, positive linear functional on the
algebra and in our cases we have:
\begin{itemize}
\item A state $\rho$ on $M_n$ is identified with a statistical operator,
i.e.\ $\rho\in \mathcal{S}_n$, and $\rho$ applied to an element $a$ of $M_n$ is given by
$\langle \rho, a\rangle = \Tr\{\rho a\}$; this is the usual quantum setup.
\item A state $p$ on $\mathcal{C}(\Omega)$ is a discrete probability density on
$\Omega$ and $\langle p, a\rangle = \sum_{\omega \in \Omega}p(\omega)a(\omega)$; this is the
classical setup.
\item A state $\Sigma$ on $\mathcal{C}(\Omega; M_n)$ is itself an element of
$\mathcal{C}(\Omega; M_n)$ such that $\Sigma(\omega)\geq 0$ and $\sum_{\omega \in
\Omega}\Tr\{\Sigma(\omega)\}=1$; the action of the state $\Sigma$ on an element $a\in
\mathcal{C}(\Omega; M_n)$ is given by $\langle \Sigma, a\rangle =\sum_{\omega \in
\Omega}\Tr\{\Sigma(\omega)a(\omega)\}$. Note the quantum/classical hybrid character of this
case.
\end{itemize}

\subsubsection{Entropies and relative entropies}
Entropies and relative entropies can be defined in very general situations \cite{OhyP93}, but
here we are interested only in the finite case, where the definitions become simpler. A finite
$C^*$-algebra $\mathcal{C}$ can always be seen as a subalgebra of a big matrix algebra $M_N$
and the definition of entropy for states on $\mathcal{C}$ is derived from the von Neumann
definition for states on $M_N$; the same type of definition applies to the relative entropy
(\cite{OhyP93}, Part I). In some sense this is the general formulation of the trick of
embedding classical probabilities into quantum states, a trick by which many results in
quantum information theory have been proved. Entropies and relative entropies are non
negative; the relative entropy can be infinite. In the case of our three $C^*$-algebras we
have:
\begin{itemize}
\item For $\rho_1,\rho_2 \in \mathcal{S}_n$, the entropy is
\begin{equation}\label{vNentr}
S(\rho_i)= - \Tr\{\rho_i \log \rho_i \}=: S_\rmq(\rho_i)
\end{equation}
(the von Neumann entropy), and  the relative entropy of $\rho_1$ with respect to $\rho_2$ is
\begin{equation}\label{qrelent}
S(\rho_1|\rho_2)=  \Tr\{\rho_1 (\log \rho_1 - \log \rho_2)\}=:S_\rmq(\rho_1|\rho_2)\,.
\end{equation}
\item In the classical case, for two states $p_1,p_2$ on $\mathcal{C}(\Omega)$, the entropy is
\begin{equation}
S(p_i) = -\sum_{\omega \in \Omega}p_i(\omega)\log p_i(\omega)=:S_\rmc(p_i)
\end{equation}
(the Shannon information), and the relative entropy is
\begin{equation}\label{crelent}
S(p_1|p_2) = \sum_{\omega \in \Omega}p_1(\omega)\log \frac {p_1(\omega)}{p_2(\omega)}=:
S_\rmc(p_1|p_2)
\end{equation}
(the Kullback-Leibler informational divergence).
\item For two states $\Sigma_1,\Sigma_2$ on $\mathcal{C}(\Omega; M_n)$ we have
\begin{gather}\label{---}
S(\Sigma_i) = - \sum_{\omega \in \Omega}\Tr
\left\{\Sigma_i(\omega)\log\Sigma_i(\omega)\right\} = S_\rmc(p_i)+ \sum_{\omega \in
\Omega}p_i(\omega)S_\rmq\big( \hat \Sigma_i(\omega)\big),
\\
\begin{split}\label{hybrelent}
S(\Sigma_1|\Sigma_2) = &\sum_{\omega \in \Omega}\Tr \left\{\Sigma_1(\omega)
\left(\log\Sigma_1(\omega)-\log\Sigma_2(\omega)\right)\right\}
\\
{}&= S_\rmc(p_1|p_2)+ \sum_{\omega \in \Omega}p_1(\omega)S_\rmq\big( \hat \Sigma_1(\omega)
\big|\hat \Sigma_2(\omega)\big),
\end{split}
\\
p_i(\omega):=\Tr \left\{\Sigma_i(\omega)\right\}, \qquad  \hat \Sigma_i(\omega):= \frac
{\Sigma_i(\omega)}{p_i(\omega)}\,.
\end{gather}
In both equations (\ref{---}) and (\ref{hybrelent}) the first step is by definition and the
second one by simple computations.
\end{itemize}

\subsubsection{Channels}
\begin{itemize}
\item A (quantum) \emph{channel} $\Lambda$ (\cite{OhyP93} p.\ 137), or dynamical map, or
stochastic map is a completely positive linear map, which transforms states into states;
usually the definition is given for its adjoint or transpose $\Lambda^*$. Let $\mathcal{C}_1$
and $\mathcal{C}_2$ be two finite $C^*$-algebras (but the definition can be extended easily)
and consider a linear map $\Lambda$  from $\mathcal{C}_1$ to $\mathcal{C}_2$ and its transpose
$\Lambda^*$ from $\mathcal{C}_2$ to $\mathcal{C}_1$, $\langle \Lambda[\Sigma], a
\rangle_{\mathcal{C}_2}=  \langle \Sigma, \Lambda^*[a] \rangle_{\mathcal{C}_1}$ for all states
$\Sigma$ on $\mathcal{C}_1$ and all $a\in\mathcal{C}_2$. If  $\Lambda^*$ is a completely
positive, unital (i.e.\ identity preserving) linear map, then both $\Lambda$ and $\Lambda^*$
are called a \emph{channel transformation} (on the states and on the algebra, respectively).
The composition of channels gives again a channel. Channels are usually introduced to describe
noisy quantum evolutions, but we shall see that also an instrument can be identified with a
channel.

\item The fundamental \emph{Uhlmann's monotonicity theorem} says that channels decrease the
relative entropy (\cite{OhyP93}, Theor.\ 1.5 p.\ 21): let $\Lambda^*: \mathcal{C}_2 \to
\mathcal{C}_1$ be a channel between finite $C^*$-algebras and let $\Lambda$ be its transpose
on states; for any two states $\Sigma, \Psi$ on $\mathcal{C}_1$, the inequality $S(\Sigma|
\Psi)\geq S(\Lambda[\Sigma]| \Lambda[\Psi])$ holds.

\item If we have three algebras $\mathcal{A},\mathcal{C}_1,\mathcal{C}_2$ and three channels
$\Lambda_1^*: \mathcal{C}_1 \to \mathcal{A}$, $\Lambda_2^*: \mathcal{C}_2 \to \mathcal{A}$,
$\Phi^*: \mathcal{C}_2 \to \mathcal{C}_1$, such that $\Lambda_1^*\circ \Phi^*=\Lambda_2^*$, we
say that the channel $\Lambda_1^*$ is a \emph{refinement} of $\Lambda_2^*$ or that
$\Lambda_2^*$ is a \emph{coarse graining} of $\Lambda_1^*$ (\cite{OhyP93} p.\ 138). In this
case, for any two states $\Sigma, \Psi$ on $\mathcal{A}$, we have $S(\Sigma| \Psi)\geq
S(\Lambda_1[\Sigma]| \Lambda_1[\Psi])\geq S(\Lambda_2[\Sigma]| \Lambda_2[\Psi])$.
\end{itemize}

\subsection{\label{subsec:channel}Instruments, channels and inequalities on relative entropies}
\subsubsection{The instrument as a channel}
Let us define the linear map
\begin{equation}\label{channI}
\Lambda_{\mathcal{I}} : M_n \to \mathcal{C}(\Omega; M_n)\,, \qquad \rho \mapsto
\Lambda_{\mathcal{I}}[\rho]\,, \qquad \Lambda_{\mathcal{I}}[\rho](\omega)=\mathcal{V}(
\omega)[\rho]\,.
\end{equation}
If $\rho \in \mathcal{S}_n$, then $\Lambda_{\mathcal{I}}[\rho]$ is a state on
$\mathcal{C}(\Omega; M_n)$; moreover, by the structure of $\mathcal{V}( \omega)$,
$\Lambda_{\mathcal{I}}$ turns out to be completely positive. The transposed map
$\Lambda_{\mathcal{I}}^*$ turns out to be
\begin{subequations}
\begin{gather}
\Lambda_{\mathcal{I}}^* :  \mathcal{C}(\Omega; M_n)\to M_n \,, \qquad a \mapsto
\Lambda_{\mathcal{I}}^*[a]\,,
\\
\Lambda_{\mathcal{I}}^*[a]=\sum_{\omega\in\Omega} \mathcal{V}(
\omega)^*[a(\omega)]=\sum_{\omega\in\Omega}\sum_{k\in K} V_k^{\omega \dagger} a(\omega)
V_k^{\omega}\,,
\end{gather}
\end{subequations}
and it is easy to see that it is a completely positive, unital, linear map: it is the channel
associated with the instrument $\mathcal{I}$.

By Uhlmann's monotonicity theorem, we have for any two states $\rho$ and $\phi$ on $M_n$
\begin{equation}\label{Uhll1}
S(\rho|\phi)\geq S(\Lambda_{\mathcal{I}}[\rho]|\Lambda_{\mathcal{I}}[\phi])\,.
\end{equation}
By eqs.\ (\ref{qrelent}), (\ref{crelent}), (\ref{hybrelent}), (\ref{channI}), (\ref{probI}),
(\ref{apostI}), inequality (\ref{Uhll1}) becomes
\begin{equation}\label{!!!}
S_\rmq(\rho|\phi)\geq S_\rmc(p_\rho|p_\phi)+ \sum_{\omega\in \Omega} p_\rho(\omega)S_\rmq\big(
\hat \rho(\omega;\rho)\big| \hat \rho(\omega;\phi)\big).
\end{equation}
This is a fundamental inequality. A possible interpretation is that the ``quantum
information'' $S_\rmq(\rho|\phi)$ contained in the couple of quantum states $\rho$ and $\phi$
is not less than the sum of the classical information $S_\rmc(p_\rho|p_\phi)$ extracted by the
measurement and of the mean ``quantum information'' $\sum_{\omega\in \Omega}
p_\rho(\omega)S_\rmq\big( \hat \rho(\omega;\rho)\big| \hat \rho(\omega;\phi)\big)$ left in the
a posteriori states.

\subsubsection{Coarse grainings}
\paragraph*{The POV measure as a channel.}
In \cite{OhyP93}, pp.\ 137-138, another channel is introduced, which involves only the POV
measure, by
\begin{equation}
\Lambda_{E}^* :  \mathcal{C}(\Omega)\to M_n \,, \qquad \Lambda_{E}^*[f]=\sum_{\omega\in\Omega}
f(\omega) E(\omega)\,, \qquad \Lambda_{E}[\rho](\omega)=p_\rho(\omega)\,;
\end{equation}
it is easy to check all the properties which define a channel. Uhlmann's monotonicity theorem
applied to this case gives the inequality (\cite{OhyP93}, pp.\ 9, 151)
\begin{equation}\label{inOP}
S_\rmq(\rho|\phi)\geq S_\rmc(p_\rho|p_\phi)\,,
\end{equation}
which is weaker than (\ref{!!!}). Indeed, inequality (\ref{!!!}) has been obtained by using a
refinement $\Lambda_{\mathcal{I}}$ of the Ohya-Petz channel $\Lambda_{E}$. To prove this
statement it is enough to introduce the map
\begin{equation}\label{Phic}
\Phi^*_\rmc: \mathcal{C}(\Omega)\to \mathcal{C}(\Omega;M_n)\,, \qquad
\Phi^*_\rmc[f](\omega)=f(\omega ) \openone\,, \qquad \Phi_\rmc[\Sigma](\omega)=\Tr\{
\Sigma(\omega)\}
\end{equation}
($\openone$ is the unit element of $M_n$); in some sense, $\Phi_\rmc$ extracts the classical
part of the state $\Sigma$. Then, it is easy to check that $\Phi^*_\rmc$ is a channel and that
$\Lambda_{E}= \Phi_\rmc\circ \Lambda_{\mathcal{I}}$.

\paragraph*{The channel $\mathcal{I}(\Omega)$.}
Let us introduce now the channel $\Phi_\rmq$, which extracts the quantum part of a state
$\Sigma$ on $\mathcal{C}(\Omega;M_n)$:
\begin{equation}\label{qchann}
\Phi^*_\rmq: M_n\to \mathcal{C}(\Omega;M_n)\,, \qquad \Phi^*_\rmq[a](\omega)=a\,, \qquad
\Phi_\rmq[\Sigma]= \sum_{\omega\in \Omega} \Sigma(\omega)\,.
\end{equation}
By eqs.\ (\ref{qchann}), (\ref{channI}), (\ref{apriori}), we get
\begin{equation}
\Phi_\rmq\circ \Lambda_{\mathcal{I}}=\mathcal{I}(\Omega)\,;
\end{equation}
$\mathcal{I}(\Omega)$ is a channel from $M_n$ into itself, which is a coarse graining of
$\Lambda_{\mathcal{I}}$. This gives the inequality
\begin{equation}
S(\Lambda_{\mathcal{I}}[\rho]|\Lambda_{\mathcal{I}}[\phi]) \geq
S(\mathcal{I}(\Omega)[\rho]|\mathcal{I}(\Omega)[\phi])
\end{equation}
or
\begin{equation}\label{IOmegabound}
S_\rmc(p_\rho|p_\phi)+ \sum_{\omega\in \Omega} p_\rho(\omega)S_\rmq\big( \hat
\rho(\omega;\rho)\big| \hat \rho(\omega;\phi)\big) \geq
S_\rmq(\mathcal{I}(\Omega)[\rho]|\mathcal{I}(\Omega)[\phi]).
\end{equation}

\section{\label{sec:Hol}Holevo's bound and related inequalities}
In quantum communication theory often the following scenario is considered: messages are
transmitted by encoding the letters in some quantum states, which are possibly corrupted by a
quantum noisy channel; at the end of the channel the receiver attempts to decode the message
by performing measurements on the quantum system. So, one has an alphabet $A$ (we take it
finite) and the letters $\alpha \in A$ are transmitted with some a priori probabilities
$p_\rmi(\alpha)$; $p_\rmi$ is a discrete probability density on $A$.  Each letter $\alpha$ is
encoded in a quantum state and  we denote by $\rho_\rmi(\alpha)$ the state associated to the
letter $\alpha$ as it arrives to the receiver, after the passage through the transmission
channel. We call these states the letter states and we denote by $\{p_\rmi, \rho_\rmi\}$ the
ensemble of the states. In the developments of the theory an important role is played by the
\emph{initial a priori state}
\begin{equation}
\eta_\rmi= \sum_{\alpha\in A} p_\rmi(\alpha)\,  \rho_\rmi(\alpha)
\end{equation}
and by \emph{Holevo's $\chi$-quantity}
\begin{equation}\label{chi}
\chi\{p_{\rmi},\rho_\rmi\}:= \sum_{\alpha\in A} p_{\rmi}(\alpha)\,
S_\rmq(\rho_\rmi(\alpha)|\eta_\rmi)= S_\rmq(\eta_\rmi)- \sum_{\alpha\in A} p_{\rmi}(\alpha)
S_\rmq(\rho_\rmi(\alpha))\,.
\end{equation}

Let us use the instrument $\mathcal{I}$, given in Section \ref{subsec:finiteInst}, as decoding
apparatus. The conditional probability of the outcome $\omega$, given the input letter
$\alpha$, is
\begin{subequations}
\begin{equation}
p_{\rmf|\rmi}(\omega|\alpha)= \Tr\{ \mathcal{V}(\omega)[\rho_\rmi(\alpha)]\} \,;
\end{equation}
then, the joint probability of input and output, the conditional probability of the input
given the output and the marginal probability of the output are given by
\begin{gather}
p( \alpha, \omega)=p_{\rmf|\rmi}(\omega|\alpha)p_{\rmi}(\alpha)\,, \qquad
p_{\rmi|\rmf}(\alpha|\omega) = \frac{p_{\rmf|\rmi}(
\omega|\alpha)p_\rmi(\alpha)}{p_\rmf(\omega)}\,,
\\
p_{\rmf}(\omega)=\sum_\alpha p( \alpha, \omega) =\sum_\alpha p_{\rmi}(\alpha)\,\Tr\{
\mathcal{V}( \omega)[\rho_\rmi(\alpha)]\}=\Tr\{ \mathcal{V}( \omega)[\eta_\rmi]\}\,.
\end{gather}
\end{subequations}
Note that $p_{\rmi|\rmf}(\alpha|\omega)$ is well defined only when $p_\rmf(\omega)>0$ and that
it must be arbitrarily completed  when $p_\rmf(\omega)=0$.

The mean information $I_\rmc$ on the transmitted letter which can be extracted in this way is
the input/output classical mutual entropy
\begin{equation}
\begin{split}
I_\rmc := &S_\rmc ( p |p_{\rmi}\otimes p_{\rmf})= \sum_{\alpha,\omega} p(\alpha,\omega)\log
\frac{p(\alpha,\omega)}{p_\rmi(\alpha)p_\rmf(\omega)}
\\
{}&= \sum_{\alpha} p_\rmi(\alpha)\, S_\rmc(p_{\rmf|\rmi}(\bullet|\alpha)|p_\rmf)=
\sum_{\omega} p_\rmf(\omega)\, S_\rmc(p_{\rmi|\rmf}(\bullet|\omega)|p_\rmi)\,.
\end{split}
\end{equation}

\subsubsection*{Holevo's bound}

By applying the inequality (\ref{inOP}) to the states $\rho_\rmi(\alpha)$ and $\eta_\rmi$ and
then by multiplying by $p_\rmi(\alpha)$ and summing on $\alpha$, one gets Holevo's inequality
\cite{Hol73}
\begin{equation}\label{Holb}
I_\rmc \leq \chi\{p_{\rmi},\rho_\rmi\}\,.
\end{equation}
In the case of a general Hilbert space, general POV measure, general alphabet, this inequality
has been proved, just by using the channel $\Lambda_E$, by Yuen and Ozawa in \cite{YueO93}.

\subsubsection*{The bound of Schumacher, Westmoreland, Wootters}

Let us consider now the a posteriori states
\begin{subequations}
\begin{gather}
\rho(\alpha,\omega):= \hat \rho\big(\omega;\rho_\rmi(\alpha)\big)=
\frac{\mathcal{V}(\omega)[\rho_\rmi(\alpha)]}{p_{\rmf|\rmi}(\omega|\alpha)} \,,
\\
\rho_\rmf(\omega) := \hat \rho\big(\omega;\eta_\rmi\big)=
\frac{\mathcal{V}(\omega)[\eta_\rmi]}{p_{\rmf}(\omega)}= \sum_\alpha p_{\rmi|\rmf}(
\alpha|\omega) \,\rho(\alpha,\omega)\,.
\end{gather}
\end{subequations}
By applying the inequality (\ref{!!!}) to the states $\rho_\rmi(\alpha)$ and $\eta_\rmi$ and
then by multiplying by $p_\rmi(\alpha)$ and summing on $\alpha$, one gets
\begin{equation}\label{SWW}
\chi\{p_{\rmi},\rho_\rmi\} \geq I_\rmc +\sum_{\omega} p_{\rmf}(\omega) \,
\chi\{p_{\rmi|\rmf}(\bullet|\omega),\rho(\bullet,\omega)\}\,.
\end{equation}
Any $\chi$-quantity introduced in this paper is defined analogously to Definition (\ref{chi});
for instance, here we have
\begin{equation}
\begin{split}
\chi\{p_{\rmi|\rmf}(\bullet|\omega),\rho(\bullet,\omega)\}:= &\sum_\alpha
p_{\rmi|\rmf}(\alpha|\omega)\, S_\rmq\big(\rho(\alpha,\omega)|\hat
\rho(\omega;\eta_\rmi)\big),
\\
\hat \rho(\omega;\eta_\rmi)=  &\sum_\alpha p_{\rmi|\rmf}(\alpha|\omega)\rho(\alpha,\omega).
\end{split}
\end{equation}

Note that $\sum_{\omega} p_{\rmf}(\omega) \,
\chi\{p_{\rmi|\rmf}(\bullet|\omega),\rho(\bullet,\omega)\}$ is the mean $\chi$-quantity left
in the a posteriori states by the instrument. Inequality (\ref{SWW}) gives an upper bound on
$I_\rmc$ stronger than (\ref{Holb}); indeed, the extra term vanishes when
$\rho(\alpha,\omega)$ is almost surely independent from $\alpha$, as in the case of a von
Neumann complete measurement, but for a generic instrument it is positive. Our proof, which is
different from the original one of \cite{SchWW96}, is based on the inequalities (\ref{Uhll1})
and (\ref{!!!}), which are new, and can be generalized to the not finite case (similarly to
the generalization of Holevo's bound in \cite{YueO93}).

\subsubsection*{A lower bound}

Let us introduce the a priori final states
\begin{subequations}
\begin{gather}
\eta^\alpha_\rmf
=\mathcal{I}(\Omega)[\rho_\rmi(\alpha)]=\sum_{\omega}\mathcal{V}(\omega)[\rho_\rmi(\alpha)]=\sum_{\omega}
p_{\rmf|\rmi}(\omega|\alpha)\, \rho(\alpha,\omega)
\\
\eta_\rmf=\mathcal{I}(\Omega)[\eta_\rmi] =\sum_{\omega}\mathcal{V}(\omega)[\eta_\rmi] =
\sum_{\omega} p_{\rmf}(\omega)\, \rho_\rmf(\omega)=\sum_{\alpha, \omega} p( \alpha, \omega) \,
\rho(\alpha,\omega) = \sum_{\alpha }p_\rmi(\alpha) \, \eta^\alpha_\rmf
\end{gather}
\end{subequations}
Similarly to the previous cases, inequality (\ref{IOmegabound}) gives
\begin{equation}\label{aprioribound}
I_\rmc +\sum_{\omega} p_{\rmf}(\omega) \,
\chi\{p_{\rmi|\rmf}(\bullet|\omega),\rho(\bullet,\omega)\}\geq
\chi\{p_{\rmi},\eta_\rmf^\bullet\}\,.
\end{equation}

\subsubsection*{Scutaru's lower bound}

In \cite{Scu95} a ``classical$\to$quantum'' channel $\Psi$ is introduced which maps states on
$\mathcal{C}(A)$ (discrete probability densities on $A$) into states on $M_n$; such a channel
is built by using the letter states $\rho_\rmi(\alpha)$. If $h$ is any discrete probability
density on $A$ and $b$ any element of $M_n$, then
\begin{equation}
\Psi^* : M_n \to \mathcal{C}(A)\,, \qquad \Psi^*[b](\alpha)= \Tr\{b\rho_\rmi(\alpha)\}\,,
\qquad \Psi[h]= \sum_\alpha h(\alpha) \rho_\rmi(\alpha)\,.
\end{equation}
By this definition we have
\begin{equation}
\Psi[p_\rmi]=\eta_\rmi\,, \qquad \Psi[p_{\rmi|\rmf}(\bullet|\omega)]= \tau(\omega)\,,
\end{equation}
where we have introduced the new family of states
\begin{equation}
\tau(\omega)= \sum_\alpha p_{\rmi|\rmf}(\alpha|\omega) \rho_\rmi(\alpha)\,;
\end{equation}
note that
\begin{equation}
\sum_\omega p_\rmf(\omega)\tau(\omega)= \eta_\rmi\,.
\end{equation}

By Uhlmann's monotonicity theorem we get
\begin{equation}\label{scu1}
S_\rmc(p_{\rmi|\rmf}(\bullet|\omega)|p_\rmi)\geq S_\rmq(\tau(\omega)|\eta_\rmi)\,;
\end{equation}
by multiplying by $p_\rmf(\omega)$ and summing on $\omega$ we get a lower bound for the
classical mutual information $I_\rmc$:
\begin{equation}\label{scu2}
I_\rmc\geq \chi\{p_{\rmf},\tau\}\,.
\end{equation}
Scutaru has proved this bound also in the non finite case.

\subsubsection*{Refinements of Scutaru's channel}

Inequalities stronger than (\ref{scu1}), (\ref{scu2}) can be obtained by finding refinements
of Scutaru's channel. There is not a canonical way to do this, but different choices are
possible; here we give only one example, inspired by the notion of compound state introduced
by Ohya (see \cite{OhyP93} pp.\ 33--34).

Let us consider two copies of $M_n$ (the ``initial'' and the ``final'' algebra) and a channel
$\Gamma$ which transforms states on $\mathcal{C}(A)$ into states on $M_n\otimes M_n$: for any
$B\in M_n\otimes M_n$ and any discrete probability density $h$ on $A$,
\begin{equation}
\begin{split}
\Gamma^* : M_n\otimes M_n \to \mathcal{C}(A)\,, \qquad &\Gamma^*[B] = \Tr_{\rmi \rmf}\{B
(\rho_\rmi (\alpha) \otimes \eta_\rmf^\alpha)\}\,, \\ &\Gamma[h] = \sum_\alpha h(\alpha)
(\rho_\rmi (\alpha) \otimes \eta_\rmf^\alpha)\,.
\end{split}
\end{equation}
Let us consider now the partial trace $\Tr_\rmf$ on the second copy of $M_n$; to take a
partial trace is a channelling transformation (similar to (\ref{Phic})) and, so,
$\Phi_\rmi=\Tr_\rmf$ is a channel (given on the states). Then, it is easy to check that $\Psi=
\Phi_\rmi\circ \Gamma$ and that $\Gamma$ is indeed a refinement of $\Psi$.

By introducing the new states
\begin{subequations}
\begin{gather}
\epsilon_{\rmi \rmf}(\omega)=\sum_\alpha p_{\rmi|\rmf}(\alpha|\omega) (\rho_\rmi (\alpha)
\otimes \eta_\rmf^\alpha)=\Gamma[p_{\rmi|\rmf}(\bullet|\omega) ]\,,
\\
\eta_{\rmi \rmf}=\sum_\omega p_\rmf(\omega) \epsilon_{\rmi \rmf}(\omega)= \sum_\alpha
p_\rmi(\alpha) (\rho_\rmi (\alpha) \otimes \eta_\rmf^\alpha)=\Gamma[p_\rmi]\,,
\end{gather}
\end{subequations}
by the same steps as in the previous case and by taking into account that the new channel is a
refinement of the old one, we get the new bounds
\begin{gather}
S_\rmc(p_{\rmi|\rmf}(\bullet|\omega)|p_\rmi)\geq S_\rmq(\epsilon_{\rmi
\rmf}(\omega)|\eta_\rmi) \geq S_\rmq(\tau(\omega)|\eta_\rmi)\,,
\\
I_\rmc\geq \chi\{p_{\rmf},\epsilon_{\rmi \rmf}\} \geq \chi\{p_{\rmf},\tau\}\,.
\end{gather}

\subsubsection*{The generalized Groenewold-Lindblad inequality}

Given an instrument $\mathcal{I}$ and a statistical operator $\eta$, an interesting quantity,
which can be called the \emph{quantum information gain}, is
\begin{equation}
I_\rmq(\eta;\mathcal{I})= S_\rmq(\eta) - \sum_\omega S_\rmq\big(\hat \rho(\omega; \eta)\big)\,
p_{\eta}( \omega)\,;
\end{equation}
this is nothing but the entropy of the pre-measurement state minus the mean entropy of the a
posteriori states.

By using the expression of a $\chi$-quantity in terms of entropies and mean entropies, as in
eq.\ (\ref{chi}), one can see that inequality (\ref{SWW}) is equivalent to
\begin{equation}\label{Iqineq}
I_\rmq(\eta_\rmi;\mathcal{I}) \geq I_\rmc + \sum_\alpha p_{\rmi}(\alpha)\,
I_\rmq(\rho_\rmi(\alpha);\mathcal{I})\,.
\end{equation}
Note that, once the instrument is fixed, the l.h.s.\ of this inequality depends only on
$\eta_\rmi$, while both $I_\rmc$ and $\sum_\alpha
p_{\rmi}(\alpha)\,I_\rmq(\rho_\rmi(\alpha);\mathcal{I})$ depend on the demixture
$\{p_\rmi,\rho_\rmi\}$ of $\eta_\rmi$.

An interesting question is when the quantum information gain is positive. Groenewold has
conjectured \cite{Gro71} and Lindblad \cite{Lin72} has proved that the quantum information
gain is non negative for an instrument of the von Neumann-L\"uders type. The general case has
been settled down by Ozawa, who has introduced the a posteriori states for general instruments
in \cite{Oza85} and in \cite{Oza86} has proved the following result.

\smallskip

\noindent\textbf{Theorem.} \textsl{Let $\mathcal{I}$ be a completely positive instrument for a
quantum system living in a separable Hilbert space and with a standard Borel space as value
space. Then,}
\begin{description}
\item (a) \textsl{the instrument $\mathcal{I}$ sends any pure input state into almost surely pure a posteriori states}
\item \textsl{if and only if  }
\item (b) \textsl{$I_\rmq(\eta;\mathcal{I}) \geq 0$, for all statistical operators
$\eta$ for which $S_\rmq(\eta)<\infty$.}
\end{description}
Now the proof is an easy application of inequality (\ref{Iqineq}).

\smallskip

\noindent\textit{Proof}. To prove that (b) implies (a) is trivial; it is enough to put a pure
state $\eta$ into the definition, which gives
\[
0\leq I_\rmq(\eta;\mathcal{I})=  - \sum_\omega S_\rmq\big(\hat \rho(\omega; \eta)\big)\,
p_{\eta}( \omega)\,.
\]
This implies that the a posteriori states $\hat \rho(\omega; \eta)$ are $p_\eta$-almost surely
pure, because the von Neumann entropy vanishes only on the pure states.

To show that (a) $\Rightarrow$ (b), the non trivial part in Ozawa's proof, let $\eta_\rmi$ be
a generic state and $\{p_\rmi,\rho_\rmi\}$ be a demixture of it into pure states; then, by (a)
$I_\rmq(\rho_\rmi(\alpha);\mathcal{I})=0$ and (\ref{Iqineq}) reduces to
$I_\rmq(\eta_\rmi;\mathcal{I}) \geq I_\rmc\geq 0$, which is (b).

This proof works also in the general case. \hfill{$\square$}

\medskip
Inequality (\ref{Iqineq}) is also interesting in itself, because it gives a link between the
quantum information gain in the case of a pre-measurement state $\eta_\rmi$ and the mean
quantum information gain in the case of a demixture of $\eta_\rmi$, a link which holds true
for any kind of instrument. The amount of quantum information has been studied and its meaning
discussed also in \cite{Jac03}, where also the connections with inequality (\ref{SWW}) and
with pure state preserving instruments have been pointed out; however, in \cite{Jac03} the
above general theorem by Ozawa \cite{Oza86} is not quoted.

\section{\label{sec:instrgen}The general case}

To treat the infinite, possibly continuous case, one needs the theory of relative entropy on
von Neumann algebras and in particular the general version of Uhlmann's monotonicity theorem
(\cite{OhyP93}, Theor.\ 5.3 p.\ 79). Here we simply present the final results, without the
proofs and the full mathematical background, which however allows for a deeper understanding
of the subject.

Let $\mathcal{H}$, where QS lives, be a separable complex Hilbert space; we denote by
$\mathcal{L(H)}$ the space of the linear bounded operators on $\mathcal{H}$, by
$\mathcal{T(H)}\subset \mathcal{L(H)}$ the trace-class on $\mathcal{H}$ and by
$\mathcal{S(H)}\subset \mathcal{T(H)}$ the set of statistical operators.

Let the alphabet $A$ be a generic set equipped with a $\sigma$-algebra of subsets
$\mathcal{B}_A$. Now, the a priori probabilities are given by a probability measure $P_\rmi$
on $(A,\mathcal{B}_A)$, the letter states are a measurable family of density operators
$\rho_\rmi(\alpha)$ and the initial a priori state is given by the Bochner integral
\begin{equation}
\eta_\rmi= \int_A P_\rmi(\rmd \alpha) \, \rho_\rmi(\alpha)\,.
\end{equation}
Let the von Neumann and the quantum relative entropies  $S_\rmq$ be defined as in
(\ref{vNentr}) and (\ref{qrelent}); then, the initial Holevo's $\chi$-quantity is
\begin{equation}
\chi\{P_{\rmi},\rho_\rmi\}= \int_A P_{\rmi}(\rmd \alpha)\,
S_\rmq(\rho_\rmi(\alpha)|\eta_\rmi)= S_\rmq(\eta_\rmi)- \int_AP_{\rmi}(\rmd \alpha)\,
S_\rmq(\rho_\rmi(\alpha))\,.
\end{equation}

Let $(\Omega, \mathcal{F})$ be another measurable space; we assume both $(A, \mathcal{B}_A)$
and  $(\Omega, \mathcal{F})$ to be standard Borel spaces (this is a very weak restriction).
Let $\mathcal{I}$ be an instrument with value space $(\Omega, \mathcal{F})$
\cite{Dav76,Oza84}, i.e.\ $\mathcal{I}$ is a map from $\mathcal{F}$ into the space of the
linear bounded operators on $\mathcal{T(H)}$ such that (i) $\mathcal{I}(F)$ is completely
positive for any $F \in \mathcal{F}$, (ii) $\sum_{j}\mathcal{I}(F_{j})[\rho] = \mathcal{I}
\big( \bigcup_j F_j \big) [\rho]$ for any sequence of pairwise disjoint elements of
$\mathcal{F}$ and any $\rho$ in $\mathcal{T(H)}$ (convergence in trace norm), (iii) $\Tr
\{\mathcal{I}(\Omega)[\rho]\} = \Tr\{\rho\}$, $\forall \rho \in \mathcal{T(H)}$. Then
\cite{Oza85}, for any pre-measurement state $\rho$ there exists a family of \emph{a posteriori
states} (unique up to equivalence), i.e.\ a family of statistical operators
$\hat\rho(\omega;\rho)$, $\omega \in \Omega$, such that the function $\omega \mapsto \hat
\rho(\omega;\rho)$ is strongly measurable and, $\forall a\in \mathcal{L(H)}$, $\forall F \in
\mathcal{F}$,
\begin{gather}
\int_F \Tr\{a\hat \rho(\omega;\rho)\} P_\rho(\rmd \omega) = \Tr\{a\, \mathcal{I}(F)[\rho]\}\,,
\\
P_\rho(\rmd \omega):= \Tr\{\mathcal{I}(\rmd \omega)[\rho]\}\,.
\end{gather}

Similarly to Section \ref{sec:Hol} we introduce the probabilities
\begin{gather}
P_{\rmf|\rmi}(\rmd \omega|\alpha):= \Tr\{ \mathcal{I}(\rmd \omega)[\rho_\rmi(\alpha)]\}\,,
\qquad P(\rmd \alpha\times \rmd \omega):=P_{\rmf|\rmi}(\rmd \omega|\alpha)P_{\rmi}(\rmd
\alpha)\,,
\\
P_{\rmf}(\rmd \omega):= P(A\times \rmd \omega)=\int_A P_{\rmi}(\rmd \alpha)\, \Tr\{
\mathcal{I}(\rmd \omega)[\rho_\rmi(\alpha)]\}=\Tr\{ \mathcal{I}(\rmd \omega)[\eta_\rmi]\}\,,
\end{gather}
the a posteriori states
\begin{equation}
\rho(\alpha,\omega):= \hat \rho\big(\omega;\rho_\rmi(\alpha)\big)\,, \qquad \rho_\rmf(\omega)
:= \hat \rho(\omega;\eta_\rmi)\,,
\end{equation}
and the final a priori states
\begin{gather}
\eta^\alpha_\rmf:=\mathcal{I}(\Omega)[\rho_\rmi(\alpha)]=\int_{\Omega} P_{\rmf|\rmi}(\rmd
\omega|\alpha)\, \rho(\alpha,\omega) \,,
\\
\eta_\rmf:=\mathcal{I}(\Omega)[\eta_\rmi] = \int_{\Omega} P_{\rmf}(\rmd \omega)\,
\rho_\rmf(\omega)=\int_{A\times \Omega}P(\rmd \alpha\times \rmd \omega) \, \rho(\alpha,\omega)
= \int_{A }P_\rmi(\rmd \alpha) \, \eta^\alpha_\rmf\,.
\end{gather}
With these notations, the classical mutual information is given by
\begin{equation}
I_\rmc := S_\rmc ( P |P_{\rmi}\otimes P_{\rmf})= \int_AP_{\rmi}(\rmd \alpha)\,
S_\rmc(P_{\rmf|\rmi}(\bullet|\alpha)|P_{\rmf})\,;
\end{equation}
let us recall that the classical relative entropy for two probability measures is defined by
$S_\rmc(\mu_1|\mu_2):= \int_{\mathcal{X}} \mu_1(\rmd x) \log \frac{\mu_1(\rmd x)}{\mu_2( \rmd
x)}$. Moreover, the definition of quantum information gain is now
\begin{equation}
I_\rmq(\eta_\rmi;\mathcal{I}):= S_\rmq(\eta_\rmi) - \int_\Omega P_\rmf(\rmd \omega)\,
S_\rmq\big(\rho_\rmf(\omega)\big)\, .
\end{equation}

Finally we can state the main results; inequalities (\ref{SWW}), (\ref{aprioribound}) and
(\ref{Iqineq}) become
\begin{gather}
\chi\{P_{\rmi},\rho_\rmi\} \geq I_\rmc + \int_\Omega P_{\rmf}(\rmd \omega)\,
\chi\{P_{\rmi|\rmf}(\bullet|\omega),\rho(\bullet,\omega)\} \geq
\chi\{P_{\rmi},\eta_\rmf^\bullet\}\,,
\\
I_\rmq(\eta_\rmi;\mathcal{I}) \geq I_\rmc + \int_A P_{\rmi}(\rmd\alpha)\,
I_\rmq(\rho_\rmi(\alpha);\mathcal{I})\,.
\end{gather}
Also Scutaru's lower bound and its refinements can be stated in this more general context, but
we do not present them here, as we do not give the analog of the results of Section
\ref{sec:instrument}. The general case will be treated in full detail in a forthcoming paper.

\begin{acknowledgments}
Work supported by the \emph{European Community's Human Potential Programme} under contract
HPRN-CT-2002-00279, QP-Applications.
\end{acknowledgments}

\end{document}